\documentclass{vgtc}                          

\ifpdf
  \pdfoutput=1\relax                   
  \usepackage{graphicx}                
  \DeclareGraphicsExtensions{.pdf,.png,.jpg,.jpeg} 
\else
  \usepackage{graphicx}                
  \DeclareGraphicsExtensions{.eps}     
\fi%

\graphicspath{{figures/}{pictures/}{images/}{./}} 
\usepackage{amsmath}
\usepackage{amssymb}

\usepackage{microtype}                 
\PassOptionsToPackage{warn}{textcomp}  
\usepackage{textcomp}                  
\usepackage{mathptmx}[osf]                  
\usepackage{times}                     
\usepackage{cite}                      
\usepackage{tabu}                      
\usepackage{booktabs}                  

\usepackage{hyperref}
\usepackage{cleveref}

\onlineid{0}

\vgtccategory{Research}

\vgtcinsertpkg

\usepackage{balance}
\usepackage{graphicx}
\usepackage{framed}

\usepackage{color}
\usepackage{subcaption}
\usepackage{enumitem}
\usepackage{xspace}
\usepackage{xcolor}
\usepackage{colortbl}

\newcommand{\red}[1]{\textcolor{red}{#1}}

\newcommand{\blue}[1]{\textcolor{blue}{#1}}
\newcommand{\orange}[1]{\textcolor{orange}{#1}}

\newtheorem{obs}{Observation}
\newtheorem{conjecture}{Conjecture}
\newtheorem{ex}{Example}

\newtheorem{example}[ex]{Example}

\newlength{\listingindent}                
\usepackage[LGRgreek]{mathastext}

\DeclareFontFamily{U}{matha}{\hyphenchar\font45}
\DeclareFontShape{U}{matha}{m}{n}{
      <5> <6> <7> <8> <9> <10> gen * matha
      <10.95> matha10 <12> <14.4> <17.28> <20.74> <24.88> matha12
      }{}
\DeclareSymbolFont{matha}{U}{matha}{m}{n}

\DeclareMathSymbol{\Lt}{3}{matha}{"CE}
\DeclareMathSymbol{\Gt}{3}{matha}{"CF}

\setlist{leftmargin=*,itemsep=0pt}

\DeclareMathAlphabet{\mathit}{T1}{cmr}{m}{it}

\newcommand{\stitle}[1]{\smallskip\noindent\textbf{#1}}

\newcommand{\ewu}[1]{\red{ewu: #1}}

\title{Design-Specific Transforms In Visualization}

\author{Eugene Wu\thanks{e-mail: ewu@cs.columbia.edu}\\ %
        \scriptsize Columbia University %
\and Remco Chang\thanks{e-mail:remco@tufts.edu}\\ %
     \parbox{1.4in}{\scriptsize \centering Tufts University}}

\teaser{
  \centering
  \includegraphics[width=\linewidth]{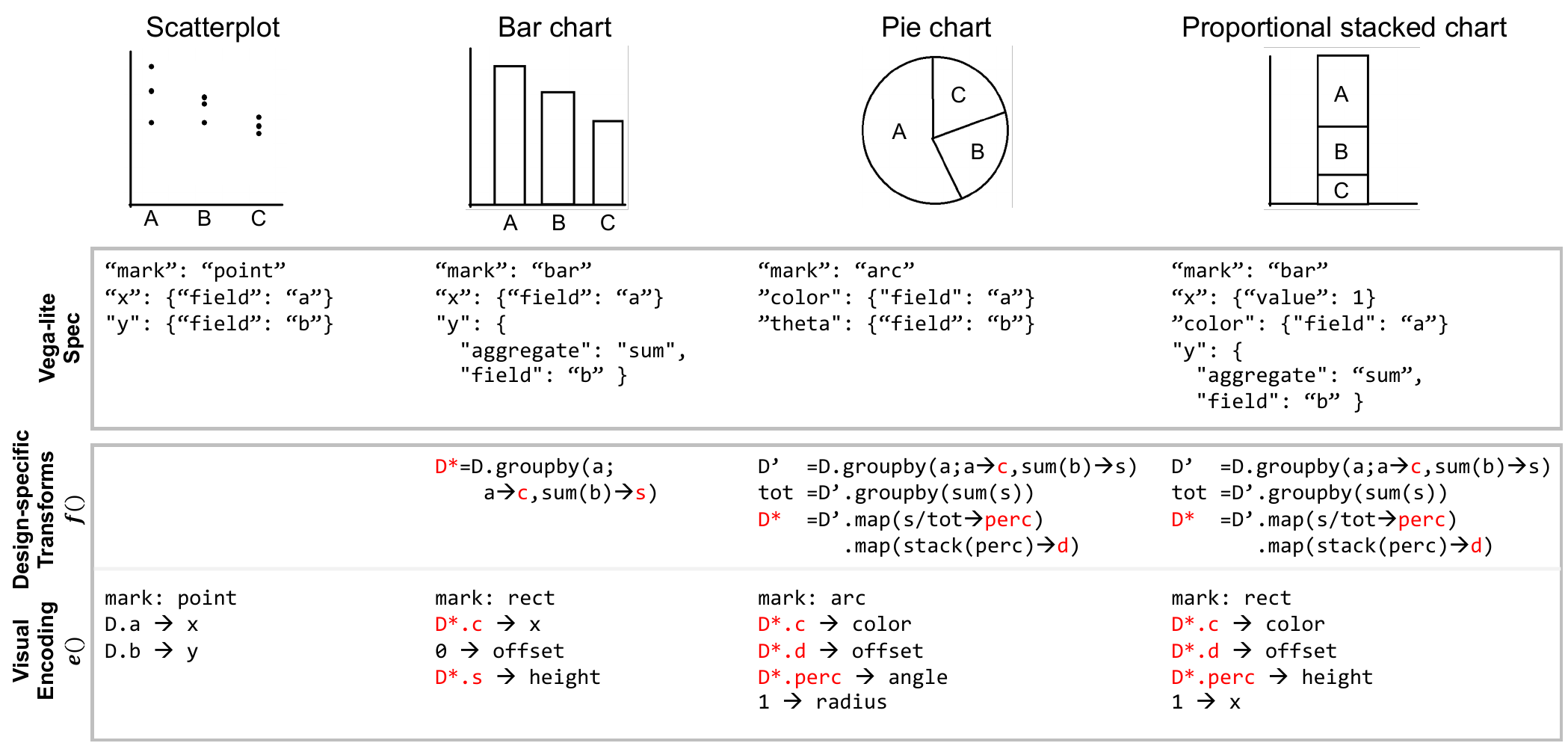}
  \caption{Existing theoretical visualization models treat the process of mapping a data table to scatterplots, bar charts, pie charts, and stacked bars as a single visual mapping step, as illustrated by the Vega-lite specifications.   However,  visual mapping crucially involves two distinct substeps: design-specific transformations that calculate statistics and layout, and visual encoding that maps data to marks.   Separating and explicitly modeling design-specific transformations helps shed light on visualizations and their relationship to other visualization designs as well as user tasks.  }
  \label{fig:teaser}
}

\abstract{In visualization, the process of transforming raw data into visually comprehensible representations is pivotal.  While existing models like the Information Visualization Reference Model describe the data-to-visual mapping process, they often overlook a crucial intermediary step: design-specific transformations.  This process, occurring after data transformation but before visual-data mapping, further derives data, such as groupings, layout, and statistics, that are essential to properly render the visualization.  In this paper, we advocate for a deeper exploration of design-specific transformations, highlighting their importance in understanding visualization properties, particularly in relation to user tasks.  We incorporate design-specific transformations into the Information Visualization Reference Model and propose a new formalism that encompasses the user task as a function over data. The resulting formalism offers three key benefits over existing visualization models: (1) describing tasks as compositions of functions, (2) enabling analysis of data transformations for visual-data mapping, and (3) empowering reasoning about visualization correctness and effectiveness.  We further discuss the potential implications of this model on visualization theory and visualization experiment design.
} 

\begin{document}


\firstsection{Introduction}
\maketitle

\begin{figure}
    \centering
    \includegraphics[width=\columnwidth]{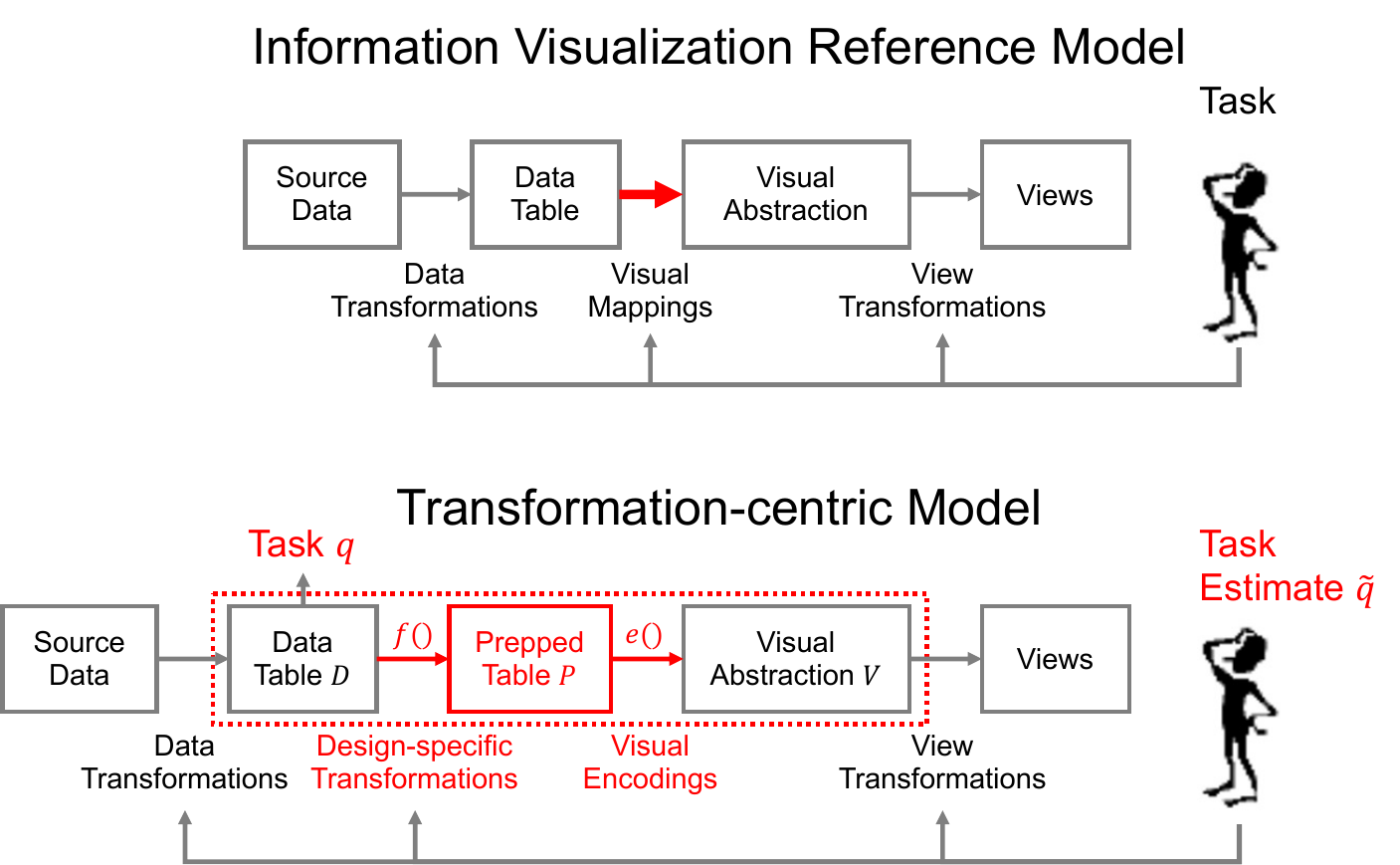}
    \caption{The proposed model in this paper differs from the Infovis Reference Model in two important ways: we decompose visual mappings into design-specific transformations (e.g., stacking, quantization, calculating statistics) from visual encoding, and we model the task $q(D)$ as a function over the input dataset that the user wishes to estimate.}
    \label{fig:computemodel}
\end{figure}

The Information Visualization Reference Model  (InfoVis Reference Model) by Card et al.~\cite{card1999readings} is a foundational framework in the visualization community towards understanding the mapping of data to visual elements.   The model (\Cref{fig:computemodel}(top)) semantically delineates the primary phases of visualization: Data Transformation cleans, reshapes, and summarizes the source dataset into data tables (for example, to derive the desired information into a spreadsheet from a large database), Visual Mapping constructs and places marks onto the spatial substrate (for example to choose between a bar chart or a pie chart), and View Transformation provides mechanisms to change the properties of visual marks and/or manipulate the viewport. 

Organizing the steps semantically has led to clean abstractions that inform how modern visualizations are developed and evaluated.   Consider the process of visualizing a pie chart. Libraries such as VegaLite, Plotly, and Matplotlib ask developers to define the data transformations used to derive the table and simply specify the visual mapping (see Figure~\ref{fig:teaser})---the library is then responsible for computing the group statistics, percentages, and mark placements needed to construct and render the marks.  
The model also helps researchers design graphical perception experiments that isolate the effects of visualization designs by focusing on the visual mapping step in isolation of data and view transformations (see a survey by Quadri and Rosen \cite{quadri2021survey}).

Despite these broad-ranging benefits, we observe that this semantic delineation obscures the process of visual mapping in a way that makes it difficult to reason about visualizations and their relationships to user tasks.   
%
Consider the four basic charts in \Cref{fig:teaser} that render a table with two numeric columns, $D(a,b)$, and how the developer specifies their visual mappings.  We choose Vega-lite~\cite{satyanarayan2016vega}  
 because it is a representative specification that is used in practice and research (e.g., \cite{2019-draco,heer2019agency,xu2020survey}).
From the specification, it would appear that scatterplot is most similar to the pie chart, as they only differ in the mark type and the visual attributes that \texttt{a} and \texttt{b} are mapped to, and the bar chart is most similar to the proportional stacked chart, as they only differ in that \texttt{a} is mapped to \texttt{x} instead of \texttt{color}.
Yet, these syntactic similarities are only superficial.  For instance, the pie chart and proportional stacked chart both encode the same information and only differ in their encoding in a circular or a rectangular layout, respectively.

We observe that the core reason for this dissonance between visualization specifications and functions is that the Visual Mapping step represents two distinct substeps.  The first involves additional transformations over the data tables in order to compute, e.g., desired statistics and spatially place marks  that are specific to the visualization design.   We term these \textbf{Design-specific Transformations}.   The second is the visual encoding that maps each row in the transformed table to a mark and data attributes to mark attributes using simple scaling functions.

Simply separating design-specific transformations and visual encodings leads to immediate insights into relationships between the four charts.  For instance, the scatterplot does not apply any additional transformations and directly encodes \texttt{a}, \texttt{b} to mark attributes. In contrast, the pie chart requires a complex sequence of design-specific transformations---it groups by \texttt{a} to derive per-group sums, computes the total sum and uses it to derive per-group percentages, and then stacks the percentage values--- before encoding the {\it derived} attributes \texttt{c}, \texttt{d}, and \texttt{perc} to mark attributes.  We can see that the stacked bar and pie chart are nearly identical and only slightly differ in their visual encodings; both are related to the bar chart, which also computes the per-group sum of \texttt{a}, but nothing further.   

This separation also sheds light on the applicability of each chart to different user tasks.  For instance, pie charts are historically a controversial chart type~\cite{tufte2001visual,few2007save,pie1,pie2,siirtola2019cost} with proponents on both sides.   Its visual encoding maps the derived \texttt{perc} and \texttt{d} (stacked \texttt{perc}) attributes.   Consistent with perception studies, we would expect the pie chart to excel at percentage estimation tasks as compared to the scatterplot and bar chart, but it is impossible to answer questions about individual input values because they have been aggregated.

Based on the above observations, this paper proposes a {\it Transform-centric Model} (\Cref{fig:computemodel}(bottom)) that extends the Infovis Reference Model in two small but important ways.  First, it expands visual mappings to explicitly model design-specific transformations and visual encoding, so that the visual abstraction $V$ is the output of applying design-specific transformations and visual encoding (simply another transformation) to the data table $D$. $V$ is then rendered and rasterized.   Second, it proposes to also model a user task as a function $q(D)$ over the input data table $D$ that the user attempts to estimate using the rendered visual abstraction.   

This separation makes it possible to analyze the relationship between different visualization designs by comparing their design-specific transformations and encodings separately.   
It also helps inform experiment design: rather than measure user responses under varying visual mappings, which confounds the impact of design-specific transformations and visual encodings, experiments can vary each in isolation for proper attribution.
Further, it models both visualization and tasks as functions over the data tables, and thus provides a way to relate the two concepts.  

The rest of this paper first introduces the transform-centric model and then presents implications for visualization theory and evaluation.

\section{Related Work}
\label{s:related}

Our work on a transformation-centric model of visualization is rooted in the rich literature on visualization theory, including work on data-visual mapping, language design, tasks, and information processing. 

Bertin's seminal work on the Semiology of Graphics lays the foundation of the theory of information visualization by introducing the concepts of data-visual mapping \cite{bertin1967semiologie}.
Many subsequent papers and applications have built on this concept, enabling efficient automatic construction of valid and informative visualizations from data (e.g., \cite{mackinlay1986automating, stolte2002polaris, mackinlay2007show, munzner2009nested, 2019-draco}, to name a few).
Modern popular visualization libraries and languages, such as D3~\cite{bostock2011d3}, Vega~\cite{satyanarayan2015reactive}, Vega-Lite~\cite{satyanarayan2016vega}, ggplot~\cite{wickham2010layered}, Mapplotlib~\cite{matplotlib}, Plotly~\cite{plotly}, and others, further facilitate the specification and manipulation of data-visual mappings, fostering the rapid development of visualizations that can be embedded in websites or data analysis environments.

While the data-visual mapping concept is powerful, it primarily addresses the properties of a visual \textit{mark}, such as its size, color, shape, etc.
It does not adequately describe visualization \textit{layout} designs and how these layouts relate to a user's task and analysis needs.
The InfoVis Reference Model by Card et al. \cite{card1999readings} bridges some of these gaps by observing that \textbf{raw data} can be processed using data transformations into \textbf{derived data}, thus allowing a user to transform their raw data into rows of information that are most relevant to their tasks. 
It then applies visual mappings to derive a visualization, and finally applies view transformations that manipulate the viewport (e.g., pan, zoom, rotate, etc).  To answer an analysis task, the user interacts with controls to manipulate the data transformations, visual mappings, or view transformations in order to achieve their desired task.  

The key limitation of this model is that it
assumes that any visualization suitable for the transformed data can be chosen interchangeably (e.g., both bar charts and pie charts are viable options for visualizing a bivariate dataset).
For designers, this is convenient, as each visualization can be described formally based on the data types that it can support.
The designer's choice of which visualization is ``best'' then depends on the user's task and general design guidelines.
For example, a designer knows when to use a pie chart because ``pie charts are more effective at visualizing part-to-whole relationships than bar charts'' \cite{eells1926relative, few2007save}. 

Although this practice represents the state of the art in visualization design (e.g., see the book by Munzner \cite{munzner2014visualization}), we observe that these general design guidelines are often subjective, difficult to replicate in empirical studies, and together lack a theoretical underpinning of reasoning about the relationship between visualizations and tasks.
Our work on a transformation-centric model aims to address these limitations.
Built on existing visualization theories and practices, we propose that by surfacing the design-specific transformations, it becomes easier for visualization designers and researchers to observe the specific transformed data being visualized and understand how these transformations can aid or limit the effectiveness of the visualization for different tasks.

\section{A Transformation-Centric Model}

\begin{figure}[b]
    \centering
    \includegraphics[width=\linewidth]{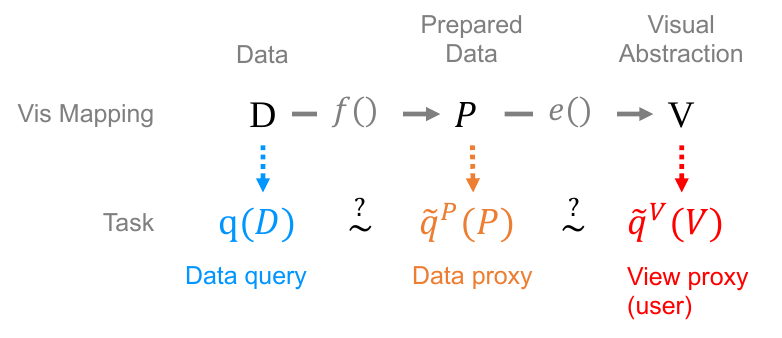}
    \caption{Summary of the Transformation-Centric Model.  An analytic task is defined as a query over $D$, and can be answered by a query over the prepared table $P$ or by the user using a rendering of the visual abstraction $V$.  }
    \label{fig:workflow}
\end{figure}

Our proposed transformation-centric model, summarized in \Cref{fig:workflow}, extends the InfoVis Reference model in two key ways (marked \red{red} in \Cref{fig:computemodel}).  The first disentangle visual mappings into explicit data transformation and visual encoding steps---this sheds light on the computational operations used to construct the visualization that goes beyond encoding and rendering marks.   The second models analytic tasks as desired queries over the input dataset that are evaluated using the visualization---this unifies visualizations and tasks under a common representation that is amenable to analysis.   

To keep the exposition simple, we will exclusively focus on the subset of the model that differs from the InfoVis Reference Model (\red{dashed red box} in \Cref{fig:computemodel}).   This is equivalent to assuming that all data transformations in the reference model are folded into the design-specific transformations, and that the view transformation is an identity function.



\subsection{The Basic Model}

We start with the output of the data transformation step in the InfoVis Reference model, a
 data table $D(a_1,\dots,a_n)$ that has $n$ attributes. 
Visual mapping is decomposed into two steps: the design-specific transformation function $f()$ produces a {\it Prepared Table} $P=f(D)$, and the subsequent encoding function $e()$  turns data rows in $P$ into marks in pixel space.  $f()$ could be described by a set of data frame operations (e.g., \Cref{fig:teaser}) or a SQL query over $D$.   For instance, $f()$ might aggregate prices by month, while $e()$ then maps month to the x-axis and aggregated prices to the y-axis of a bar chart.  The resulting {\it visual abstraction} $V=e(P)=e(f(D))$ is a table of mark ``rows'' (e.g., SVG elements) that will ultimately be rendered in the view that the user sees.

The key characteristic of this pipeline is that $f()$ encapsulates {\it all} logical transformations---filters, derivations, statistics, layout, and even per-pixel aggregation~\cite{jugel2014m4}---so that $e()$ is only responsible for constructing one mark for each row in $P$, and assigns data attributes to mark properties based on linear scaling functions from an attribute's domain to the target range in pixel space\footnote{We restrict scaling functions to be linear so that non-linear transformations like log transforms are encoded in $f()$.}.  The output of $e()$ is a mark table that can be directly rendered (e.g., using an SVG renderer) to produce an image.   Under this model, all transformations related to the visualization are made explicit, and encoding is simply another transformation operation that applies simple scaling functions to data attributes.   

\begin{example}
    \Cref{fig:teaser} depicts design-specific transformations $f()$ and encodings $e()$ for four common visualization types.   The scatterplot uses an identity function for $f()$. The bar chart sums $b$ by $a$, while the pie and stacked charts additionally compute percentages for each group and sum the percentages.   It is also immediately clear which visualizations differ only in visual encoding (pie and stacked charts), and which also differ in their design-specific transformations (scatter, bar, and the rest).   
\end{example}

\subsection{Analytic Tasks as Proxy Queries}
Let us now consider a subset of analytic tasks that can be formulated as computations over the input data $D$.  
The user's task is to answer a question by calculating the desired value(s) or decision, expressed as a function or {\it query} $\blue{q(D)}$ over the source data $D$. These include low-level tasks in popular taxonomies~\cite{amar2005low} (e.g., identifying trends, finding outliers, value reading) as well as the query and manipulate task types in Brehmer and Munzner's task typology~\cite{brehmer2013multi}. 

Given only access to the visualization, the user attempts to answer $\blue{q(D)}$ by formulating a strategy that involves some combination of reading data values in the visualization and mental calculation.  We model this as a \textbf{view-level proxy query} $\red{\tilde{q}^V(V)}$ (or view proxy for short) over the visual abstraction.  We use the tilde above the $q$ to denote that it may be an estimate, and we use the comparison symbol $\stackrel{?}{\sim}$ to denote that a strategy may not exist.
\begin{align*}
    \blue{q(D)} 
    &\stackrel{?}{\sim} \red{\tilde{q}^V(V)} = \red{\tilde{q}^V(e(f(D)))}
\end{align*}
\begin{example}\label{ex:viewproxy}
The user wants to compare the ratio of $A$ versus $B$ using the proportional stacked chart in \Cref{fig:teaser}.   One possible view proxy $\red{\tilde{q}^V}$ is to estimate the length of $A$ and $B$ and perform mental calculations to estimate their ratio.   Talbot et al.~\cite{talbot2014four} suggest that users often compute the ratio $\frac{A}{A+B}$, which leads to a less accurate estimate.
\end{example}
As suggested in the preceding example, a major challenge with $\red{\tilde{q}^V}$ is that it is defined over the rendered visualization and relies on discerning the user's strategy.  However, graphical perception and cognitive studies only provide a partial understanding because they measure properties of $\red{\tilde{q}^V}$ as a black box (e.g., latency, accuracy, enjoyment) rather than exposing the individual steps that makeup $\red{\tilde{q}^V}$.  Think-aloud protocols ask users to describe their thought process as a way to surface their strategies and estimate $\red{\tilde{q}^V}$, but they are time-consuming and expensive, and the validity of think-aloud protocols has been questioned~\cite{alhadreti2018rethinking,ramey2006does}.

A benefit of formulating visualization as a sequence of design-specific and encoding transformations and reducing the complexity of the encoding is that we can analyze the step immediately before encoding and rendering---the prepared data.  This allows us to disentangle encoding from computation by checking whether a strategy exists given $P$.  Since $P$ is a table, the \textbf{data-level proxy query} $\orange{\tilde{q}^P(P)}$ (or data proxy for short) can be expressed as a SQL query over $P$ if it exists.  Further, since it is independent of the user, $\orange{\tilde{q}^P}$ could be automatically derived from $q,f,e$ by using standard query rewriting techniques from data management~\cite{halevy2001answering}.
\begin{align*}
    \blue{q(D)}     
    &\stackrel{?}{\sim} \orange{\tilde{q}^P(P)} = \orange{\tilde{q}^P(f(D))}
\end{align*}

\begin{example}\label{ex:trade-off}
    Let us estimate the total percentage of groups $A$ and $B$ in \Cref{fig:teaser}.  
    Since $f(D)=D$ for the scatterplot, its $\tilde{q}^P$ would need to compute the sum of $b$ per $a$, translate the sums into percentages, and then add the percentages for $A$ and $B$.
    The $\tilde{q}^P$ for the bar chart needs to compute and sum the percentages, while using the pie and proportional stacked bar chart only requires summing two values.
    
    
\end{example}

The notion of a data-level proxy query is useful because its existence serves as an ``upper-bound'' on what the user will be capable of answering from the visualization.  This is because $q$ is a function of only $D$, $V$ does not contain more information than $P$ because the encoding step $e()$ is a trivial mapping, and there is no bound to the complexity of $\orange{\tilde{q}^P}$.   Thus, if it is not possible to compute $q$ using $P$ ($\orange{\tilde{q}^P}$ does not exist), then it is not possible to answer $q$ using $V$ and thus,  $V$ is {\it inappropriate} for  $q$.

\begin{example}\label{ex:upperbound}
  The user wants to know the number of rows per value of $a$ in \Cref{fig:teaser}, which can be expressed as $q=$\texttt{D.groupby(a;count()$\to$c)}.   Given the outputs of the design-specific transforms, we see that $\orange{\tilde{q}^P}$ is only defined for the scatter plot because the other visualizations group by $a$ but only compute the sum in each group.  Thus, they are inappropriate for $q$.
\end{example}

\begin{figure}
    \centering
    \includegraphics[width=.85\linewidth]{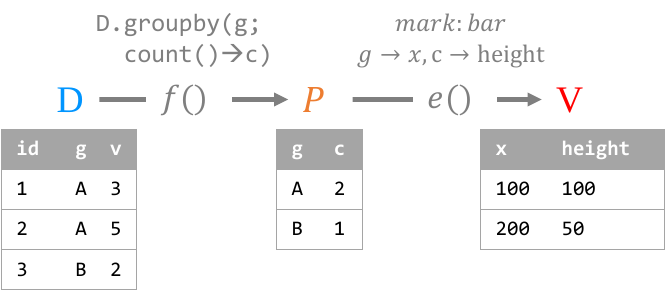}
    \caption{Point marks that render counts by $g$. }
    \label{fig:workflowex}
\end{figure}

\begin{example}\label{ex:upperbound}
\Cref{fig:workflowex} is a scatterplot of counts by $g$.   If the task is to estimate the count of group $A$, the data proxy would filter using $g=`A\textrm'$ and read the value of attribute $c$, while the view proxy might find the point for $A$ and read its value from the mark's y-position.  If the task is to estimate the average $v$ in each group $g$, $P$ cannot compute the statistic, so no such data proxy nor view proxy can exist.
\end{example}

\subsection{Simplifying the InfoVis Reference Model}

The InfoVis Reference Model in \Cref{fig:computemodel} starts with data transformations of the source data to derive a data table before applying visual mappings (we refer to this transformation as $t()$), while the Transform-Centric Model deconstructs the visual mappings into design-specific data transformations $f()$ followed by visual encoding, $e()$.  
Since $t()$ and $f()$ are both transformations on data, we can ``fold'' $f()$ into $t()$.  This results in the original Reference Model with the change that the source data derives the prepared table, and that visual mappings are now well-defined encodings.
In this sense, this paper has not increased the complexity of the original InfoVis Reference Model, but rather clarifies the semantics of the visual mappings and the relationship between visualizations and analytic tasks.

\if{0}
Next, we will first present two perspectives on this model and then illustrate how this model helps reason about visualization correctness, effectiveness, and experimental design.   
\ewu{Mention that we will be presenting definitions, observations, and claims and what they each mean.}
The discussion will be in the context of two candidate visualizations $V_1=e_1(f_1(D))$ and $V_2=e_2(f_2(D))$ and a user task $q(D)$ (\Cref{fig:v1v2}).   The task is potentially answerable by $V_i$ via $\tilde{q}^V_i(V_i)$, and by the transformed datasets via the proxy $\tilde{q}^P_i(P_i)$.  To keep the notation simple, we refer to $q$, $\tilde{q}^V_i$, and $\tilde{q}^P_i$ without their arguments when it is clear.

\begin{figure}[h]
    \centering
    \includegraphics[width=.6\columnwidth]{figs/v1v2.pdf}
    \caption{Diagram used for model-based analyses.  The subscript ${}_i$ is used to denote both proxy queries for each prepped dataset $P_i$ and visual abstraction $V_i$.   }
    \label{fig:v1v2}
\end{figure}
\fi

\section{Implications of the Model}

We now present several implications and potential applications of the transform-centric model. 

%

\subsection{The No Free Lunch Conjecture}

Assuming the view-level proxy query $\tilde{q}^V$ exists, we can reframe its earlier definition in terms of function composition to see how the user task $q$ decomposes into the view proxy $\tilde{q}^V$, encoding $e()$, and design-specific transformation $f()$ steps:  
$q \sim \tilde{q}^V\cdot e \cdot f$
Since the complexity of $e()$ is fixed, the work to compute the task is necessarily delegated to the visualization (through $f()$) or to the user (through $\tilde{q}^V$).  This leads us to the following conjecture:

\begin{conjecture}[No Free Lunch]
For any given task, either the human or the computer has to perform the necessary computation. 
In general, humans prefer to engage the ``fast brain'' and avoid unnecessary mental tasks\cite{daniel2017thinking}. 
As a result, humans will tend to consider a visualization as ``effective'', ``easy to use'', or ``good'' if the computer does the computation and the visualization directly encodes the desired information.
\end{conjecture}

At one extreme, the visualization fully pre-computes the user's task. This means the task result is visually encoded as a mark property in $V$ and, barring a poor encoding choice, can be answered via a visual lookup.  Checking this is also straightforward: suppose $q(D)$ is a scalar (e.g., the user wants to compare two statistics or estimate the slope of a trend), then the visualization has {\bf precomputed} the task if $q(D)\in f_i(D)$.   The simplest variation of this is if $f=q$, whereupon the visualization simply renders the task answer.
At the other extreme, $f()$ is the identity function and the full burden of the task is left to the user.   

\begin{example}[Pie charts and ``parts-to-whole'' relations]
Why is a pie chart better than a bar chart at visualizing ``parts-to-whole'' relations?
Suppose the task is to estimate the percentage of $A$: \texttt{q=D.filter(a=`A').groupby(sum(b))/D.groupby(sum(b))}.   
Using the pie chart in \Cref{fig:teaser} is a visual lookup because the percentage is encoded as the angle, whereas the bar chart user would need to perform mental arithmetic that divides the $A$ bar with the sum of all three bars.  
\end{example}

Between the extremes, the visualization can also pre-compute part of the task.  For instance, \Cref{ex:trade-off} describes the case where the pie chart and proportional stacked chart directly encode the percentages for $A$, $B$, and $C$, so the user only needs to estimate $A+B$.   The bar chart also partially pre-computes the sum of each group, but the user must perform the rest of the calculations.

Some types of pre-computation can also hinder the user.
One type is an adverse transform that is not only not needed for answering $q$, but forces the user to invert the transformation.  For instance, if the user wants to compare the percentages of groups $B$ and $C$ in \Cref{fig:teaser}, the proportional stacked chart's additional stacking operation causes the bars to be misaligned, which is detrimental to user judgement~\cite{talbot2014four}.
In contrast to adverse transforms, where it is still possible for the user to estimate their task (albeit with more difficulty), lossy transforms make this impossible.   Common examples include aggregating to a granularity that is too coarse, or smoothing a line when the details are important.
These two types makes clear that user task error can be due to very different reasons: one is due to additional effort/difficulty to ``invert'' adverse transforms, while the other is because the user is forced to estimate values given partial information.



We close this subsection by noting that the above examples are all based on the same four visualizations in \Cref{fig:teaser}.  Yet, each is either effective or ineffective based on the user task and how much the visualization has productively pre-computed parts of it.  This leads us to our second conjecture:

\begin{conjecture}[Visualization Utility is Task-dependent]
    Visualization ``Utility'' is orthogonal to considerations such as perceptual accuracy or cognitive load.  It is correlated to the extent that $f()$ pre-computes the user task in such a way that $\tilde{q}^V$ is as simple to visually estimate as possible.
\end{conjecture}

\subsection{Pre-computation and Task Flexibility}
\label{ss:flexibility}

Taken in isolation, the previous subsection implies that the visualization should always pre-compute the user's task because it shifts the work from the user to the visualization.    However, the no free lunch conjecture is specific to a single visualization and single task.  If we consider a  {\it set of tasks}---perhaps those that the designer expects the user is interested in---then the benefits of pre-computation are not always clear because ``specializing'' the visualization for one query in the task may come at the cost of making another query harder or impossible.

\begin{example}
A scatterplot is popular for bivariate data, and has been documented to support a large range of tasks reasonably well as compared to other bivariate visualization designs.  
However, practitioners have found scatterplots to be difficult to use for many everyday visualization consumers. Notably, the New York Times has stopped using scatterplots in their visualization designs, citing its high difficulty of use for their readers.   Both of these observations make sense because $f()$ is the identity.  The visualization can technically answer any question $D$ can (to the granularity of a pixel), but all of the work is delegated to the user.
In this regard, a scatterplot is broadly useful but burdensome because it is a ``blank slate.''

In contrast, the pie and proportional stacked charts in \Cref{fig:teaser}  greatly simplify tasks involving parts-to-whole tasks.  However, it is not useful for other tasks,  such as those involving count statistics. In this sense, these visualizations can be considered more ``task-specific''.
\end{example}

\begin{obs}[Task Flexibility vs Task Effectiveness]
Visualization design entails a trade-off between task-specific efficiency (e.g., pie chart) and flexibility (e.g., scatterplot).     When using flexible visualizations, leveraging visualization proxies as view-level strategies can reduce the perceived difficulties for some subsets of tasks.  
In either case, it is important to make the set of intended tasks explicit so that the visualization design can be properly assessed.  
\end{obs}


\begin{example}
    
To illustrate the implications of this observation, let us consider four increasingly larger sets of tasks, where each set adds to the one before it, that a designer considers before choosing a visualization (from the list of visualizations in \Cref{fig:teaser}).

\begin{itemize}
\item \textbf{(T1) Report the percentage of a specific group:} The task computes a single scalar, so the designer can compute the percentage value and report it. 
\item \textbf{(T2) Report the percentage of some groups:} Since there are only three groups $A$, $B$, $C$, the designer can again compute the statistic and report it (perhaps as an infographic).
\item \textbf{(T3) Calculate the sums or differences (of percentages) between any two groups:} The user may wish to estimate the total percentage of several groups or compare the percentages of different groups. Since enumerating the results of all tasks can be unwieldy, the designer can instead compute the percentages of each group and encode them as e.g., a pie chart, proportional stacked chart, or unstacked proportional chart.  The user is free to perform the calculations themselves, but the visualization aids them by pre-computing the percentages.
\item \textbf{(T4) Calculate the sums or differences of any statistics between any two groups:} A single static visualization is unlikely to support arbitrary statistics of the groups, and introducing interactions to choose or define the desired statistic may be helpful.  The visualization still aids the user by pre-computing the desired statistics, but the user can still perform the calculations visually.
\end{itemize}
\end{example}

%

\subsection{Where Do Tasks Come From?}

The notion of starting visualization design with a set of tasks is convenient because they are well-defined, amenable to automatic analysis, and form the basis of the concepts of view and data proxies.  On the other hand, end users rarely start with a concrete task that can be formulated as a query---so where do these tasks come from?  


Although a visualization end-user may not start with a specific query, we argue that visualization authors of new visualizations and dashboards, as well as developers of new visual analysis, exploration, and authoring systems (e.g., Tableau, Looker), implicitly design toward a potential set of tasks.    The set of such tasks can vary from a handful of queries to the set of all possible SQL queries.   

\begin{figure}
    \centering
    \includegraphics[width=0.8\linewidth]{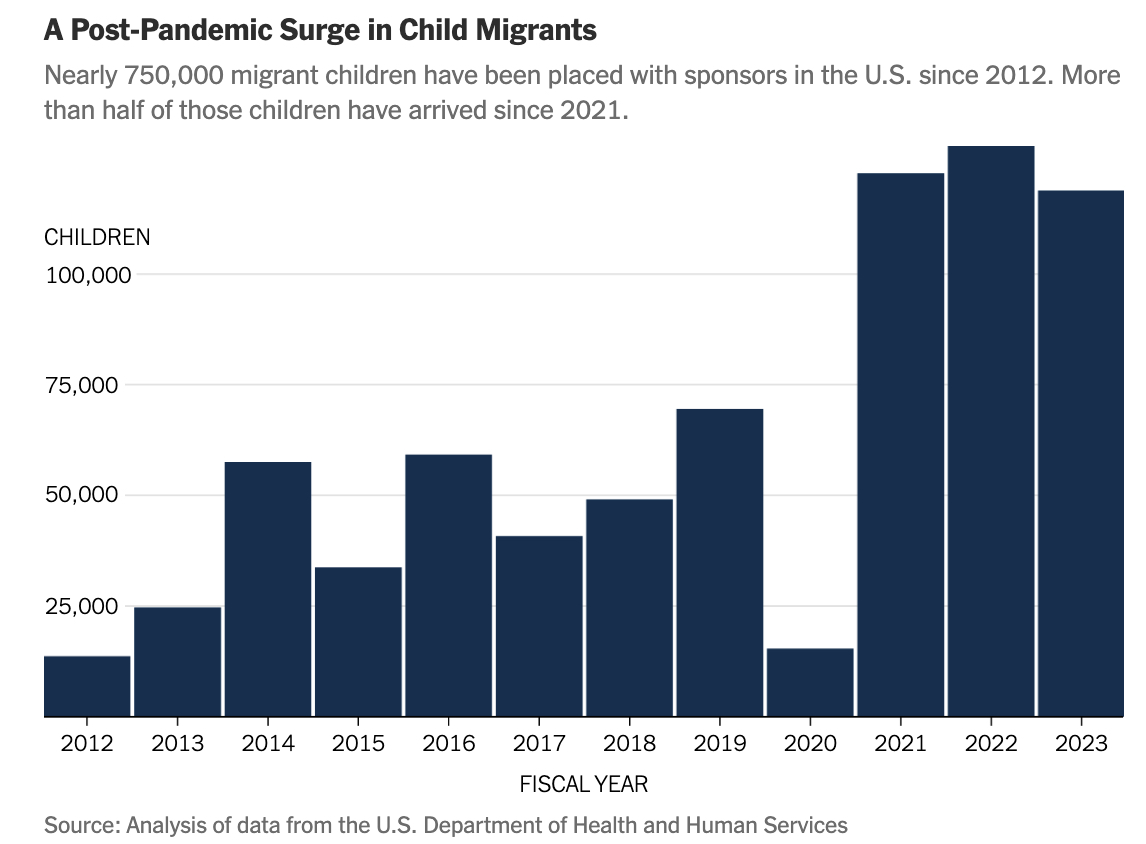}
    \caption{New York Times visualization that tells a data story about the post-pandemic surge in child migrants.}
    \label{fig:nyt}
\end{figure}

\stitle{Infographics and Data Stories} are designed with an intended message and a set of questions that the author wants the user to consider.   For instance, \Cref{fig:nyt} describes a jump in the number of child migrants after the Covid-19 pandemic (2021 and beyond).  The story and set of tasks compare the number of migrants before, the beginning, and after the start of the pandemic.  The bar chart is likely chosen to aid these comparisons.

\begin{figure}
    \centering
    \includegraphics[width=.8\linewidth]{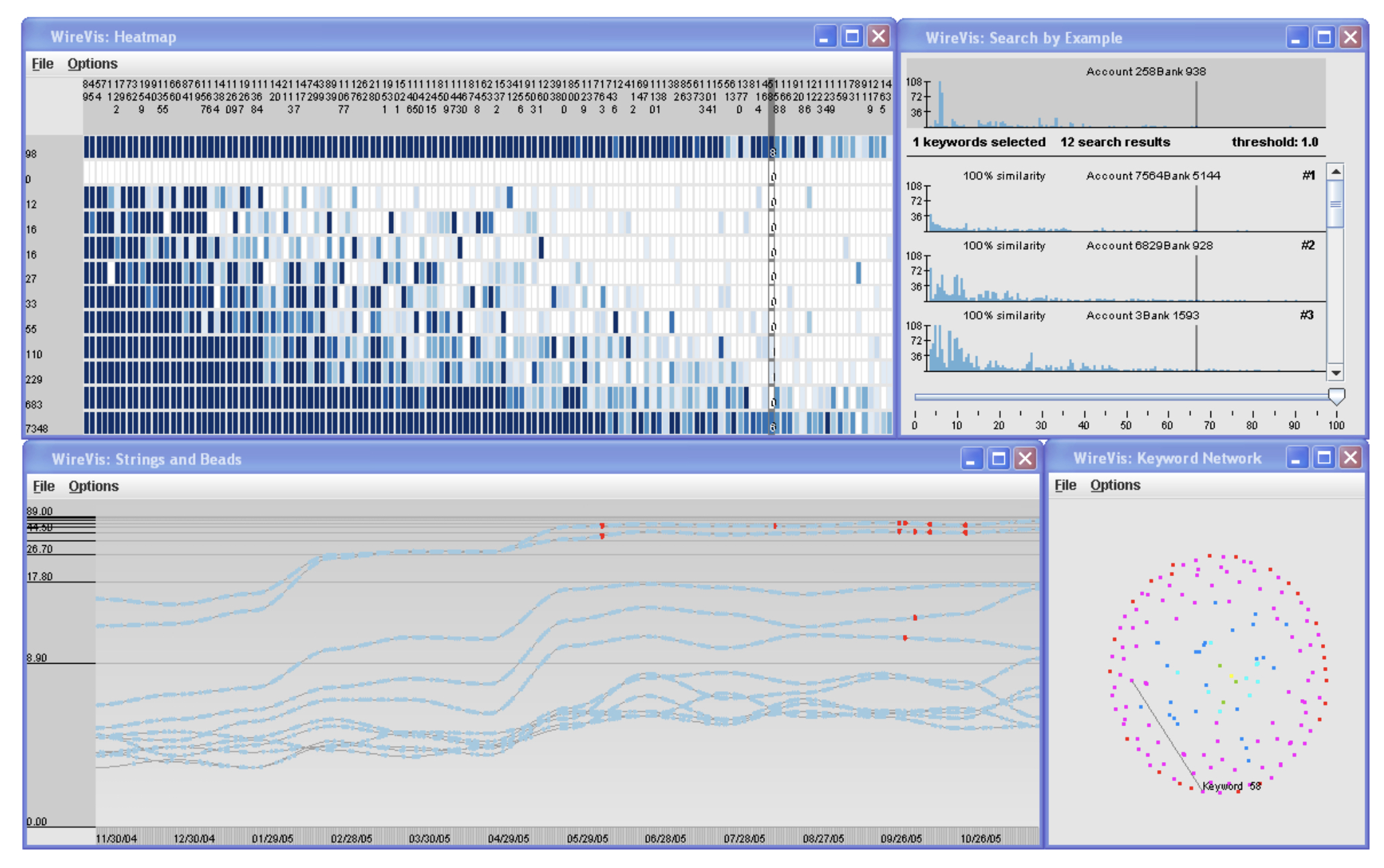}
    \caption{WireVis system to detect fraudulent banking activity. }
    \label{fig:wirevis}
\end{figure}

\begin{figure*}
    \centering
    \includegraphics[width=\textwidth]{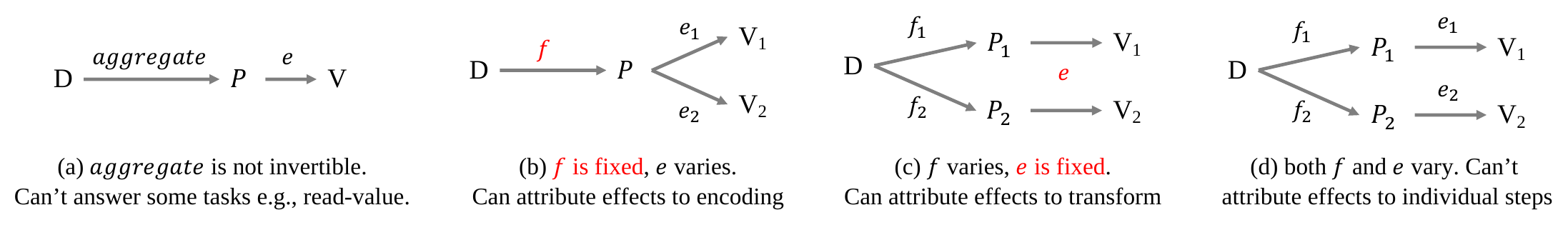}
    \caption{Application of Transformation-centric Model to experimental design.   (a) Non-invertible data transformations restrict the tasks that the visualization can answer.  Fixing one step and varying the other allows experimental effects to be attributed to (b) data transformation or (c) encoding.   (d) Varying both steps can evaluate end-to-end effects but cannot be attributed to an individual step.  }
    \label{fig:appropriateness}
\end{figure*}

\stitle{Visual Analytics Systems}  often start with need-finding studies or end-user surveys to understand the analytic tasks that the system should support.  However, these tasks are usually described informally.   For instance,  WireVis~\cite{chang2007wirevis} is a coordinated visualization system designed to help bank analysts discover fraudulent activity based on the keywords in the wire transactions (\Cref{fig:wirevis}).   The system prepares the raw transaction data by clustering accounts using the keywords in their transaction activity.   When the user searches by or chooses a keyword, the lower left view plots the amount of activity for each cluster that contains the keyword over time.   The view and intended set of tasks can be fully expressed as a parameterized SQL query that groups by cluster and date, whose filter is parameterized by keyword and cluster id.  
 Similarly, the other views are designed to support analysis functionality that can also be expressed as sets of parameterized queries. 

\stitle{Exploration Systems} let the user quickly answer a broad set of tasks.   For instance, Tableau's VizQL~\cite{stolte2002polaris} was designed to support business analytic queries, namely SQL filter group-by queries that can be expressed by a data-cube~\cite{gray1997data}.   The visual analytic interface's shelf-based interactions were then designed to ``fill in'' the grouping expressions, filters, and aggregation functions.  

\smallskip
\noindent Prior works have proposed data-driven ways to learn a set of tasks from a historical query log~\cite{chen2022pi2}, analyses executed in a Jupyter Notebook~\cite{tao2022demonstration}, or from natural language using large language models~\cite{chen2022nl2interface}.
Chen et al.~\cite{chen2023dig} propose a Data Interface Grammar to compactly represent sets of queries and describe a formal mapping from any DIG grammar to the set of valid interactive visualization interfaces.   



\section{Implications for Visualization Evaluation}


In this section, we discuss the implications of the proposed transformation-centric model for visualization evaluation. 
First, we demonstrate how this model can be used to assess the appropriateness of evaluation designs when comparing visualizations for a given task. 
Second, we suggest that applying a cost model can lead to a quantitative measurement of visualization effectiveness for specific tasks 

\subsection{What is Measured When Comparing Visualizations?}

When evaluating different visualization designs,
we should consider design-specific transformations and visual encoding as separate experimental factors.   As such, two visual mappings only make sense to compare if they differ in only one of the factors, or if all combinations are evaluated. Below we provide examples of visualization comparisons and discuss how our proposed model can be used to describe their appropriateness.

\stitle{Evaluating Data Transformations.} Consider the task ``read the value of a data point'' using a scatterplot and a bar chart.
For a scatterplot, the x- and y-values of a data point can be readily read from the visualization.
Conversely, for a bar chart, it is not possible to read the values of any individual data point because (1) the bar chart visualizes the sum of the values, and (2) the $sum$ operation is not invertible -- that is to say, a viewer cannot perform the mental arithmetic to reverse the operation and retrieve the values of each data point (see Figure~\ref{fig:appropriateness}(a)). In this sense, we state that it is \textbf{inappropriate} to compare a scatterplot and a bar chart for this task.

\stitle{Evaluating Perceptual Effectiveness.}
Consider the task ``compare two values'' using a pie chart and a proportional stacked bar chart.
The two visualizations require the same preparatory computations and visualize the same resulting derived information.
However, the two differ in their visual encodings, with a pie chart using wedges in a Polar coordinate and the proportional stacked bar chart using rectangles in a Cartesian coordinate system.
In this sense, we note that it is \textbf{appropriate} to compare a pie chart and a stacked bar chart for the given task. 
Since the preparatory computations and the derived information are the same, the comparison between the two visualizations effectively measures the relative \textbf{perceptual effectiveness} of their visualization encodings (see Figure~\ref{fig:appropriateness}(b)).

\stitle{Evaluating Mental Arithmetic.} Consider the task ``read the value of a bar'' using a bar chart and a proportional stacked bar chart.
For a bar chart, the visualization encodes the desired information, which a viewer can readily retrieve.
For a proportional stacked bar chart, assuming that the user knows the sum ($sum(s)$), the user can inverse the mapping function from proportions information back to absolute values. 
In this sense, we note that it is \textbf{appropriate} to compare a bar chart and a stacked bar chart for the given task. However, since the two visualizations use similar visualization elements, a comparison between a bar chart and a proportional stacked bar chart is effectively measuring the cost of the \textbf{mental arithmetic} to invert the sum operation (see Figure~\ref{fig:appropriateness}(c)).

\stitle{Evaluating Multiple Criteria.} Consider the task ``read the value of a bar/wedge'' as before, but instead of comparing a bar chart with a stacked bar chart, we compare a bar chart with a pie chart.
Same as before, for a bar chart, the visualization encodes the desired information, which a viewer can readily retrieve.
However, when compared a pie chart, two criteria are measured at the same time -- the cost of \textbf{mental arithmetic} and \textbf{perceptual effectiveness}.
Mental arithmetic represents the cost of inverting the mapping function necessary for a pie chart (same as a proportional stacked bar chart), and perceptual effectiveness represents the relative costs of encoding the data as rectangles versus wedges/circles.
Although this evaluation is technically \textbf{appropriate}, it conflates two criteria in the evaluation, rendering the outcome less informative (see Figure~\ref{fig:appropriateness}(d)).

\subsection{Data Proxies and Analytic Shortcuts}

The preceding examples have assessed the effectiveness of a visualization's pre-computation by describing the steps involved in answering the task from the perspective of the data-level proxy query (data proxy).  The underlying implication is that the difficulty of using the visualization is proportional to the complexity of the data proxy.   For instance, parts-to-whole calculations are a simple read operation using a pie chart but require mental arithmetic when using a scatterplot or bar chart.  Since the data proxy can be described as a sequence of primitive operations (e.g., filter, read, calculate, etc.), it may be possible to assign costs to each operation to estimate the difficulty of the data proxy, and consequently to the view-level proxy query.  Such a ``cost model'' is analogous to those used to evaluate user interfaces~\cite{gajos2004supple} and SQL queries~\cite{selinger1979access}, and could e.g., serve as a null model for evaluating visualization task-effectiveness, rank visualization designs, or evaluate the difficulty of multi-step tasks.

Of course, a key drawback of this idea is that {\it users are not computers}---they make use of visual and cognitive heuristics, as well as domain expertise and experience.  How can this be adequately modeled in a way that is useful?

\begin{figure}[t]
    \centering
    \includegraphics[width=\linewidth]{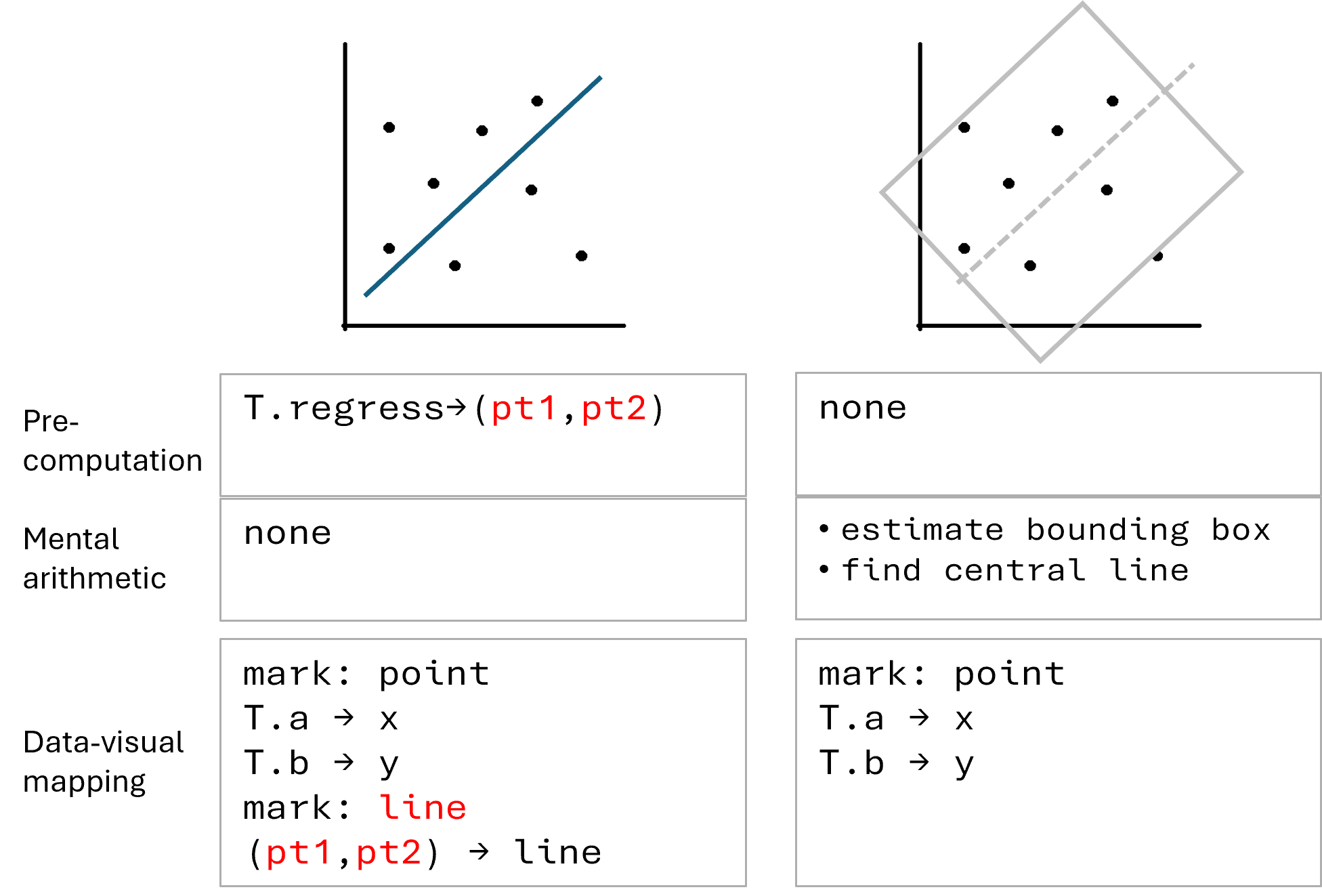}
    \caption{Two approaches to estimate the slope of the best linear-fit line.  
 (Left) the visualization pre-computes and renders the best fit line, (right) expert users use analytic shortcuts to estimate the slope.}
    \label{fig:remco-warmup2}
\end{figure}

Consider the task of estimating the slope of the regression line fit to a set of points in \Cref{fig:remco-warmup2}.   The left side pre-computes the task by fitting and rendering a line so the user can directly read the slope.  In contrast, expert visualization users may develop analytic shortcuts, such as the use of visual features as proxies to solving the complex mental arithmetic\cite{yang2018correlation, harrison2014ranking} (right side).  
Thus, although the scatterplot does not require pre-computation, the task is easy to perform by using such shortcuts \cite{braun2024beware, braun2023visual}.  

One way to model this is by extending the set of primitive operations with a library of meta-operations that are cheap for the user to evaluate and cover multiple steps in the data-level proxy query  $\tilde{q}^P$.   Different expertise levels can be modeled as different libraries of meta-operators.

\stitle{Application to Experiment Design.}
We now sketch a potential way where a cost model can be used to derive a baseline when evaluating visualization effectiveness for specific tasks.   Given a task, we can use the cost model to find the lowest cost strategy, and use that as a ``null model'' of the user's expected strategy for answering the view proxy.   
If the results of a user study coincide with the null model, then it suggests that the user is using the lowest cost strategy i.e., a rational strategy.
If the results are worse than predicted by the null model, then it suggests that the user is performing an inefficient strategy, perhaps due to the visual design.  Again, it may be possible to enumerate a handful of higher-cost strategies to design follow-up experiments to isolate the specific strategy that user uses.
Finally, if the results are better than predicted, it suggests that the user is performing analytic shortcuts that can be further investigated.


%
%
%

%
%


\section{Conclusions and Discussion}
In this paper, we present a transformation-centric model that extends the Information Visualization Reference Model (InfoVis Reference Model) by Card et al. \cite{card1999readings}. Our model separates design-specific transformations from visual encodings, offering two theoretical framings---a view-level proxy query and a data-level proxy query---for considering how transformed data and the resulting visualizations can be used to address a user's data tasks. Based on our model, we propose the ``No Free Lunch'' conjecture for visualization design and a framework based on the trade-off between pre-computation and visualization task flexibility. Finally, we argue that this model provides clearer insights into the relationships between different visualization designs based on a theoretical cost model and informs better experiment design by isolating visual encoding and transformations.

%



\bibliographystyle{abbrv-doi}

\bibliography{template}
\end{document}